\documentclass[a4paper,11pt]{article}
\pdfoutput=1

\usepackage{cancel}
\usepackage[T1]{fontenc}
\usepackage{siunitx}
\usepackage{jheppub}
\usepackage{simpler-wick}

\newcommand{\spsq}[1]{\left[#1\right]}
\newcommand{\spag}[1]{\left< #1 \right>}
\newcommand{\mix}[3]{\left<#1\right| #2 \left|#3\right]}
\newcommand{\gdf}[1]{\delta^{(#1)}\! \left(Q\right)}

\title{Soft Factor Structure of MHV Amplitudes for Massless Charged Particles}

\author{Christoph Bartsch,}\emailAdd{christoph.bartsch@mff.cuni.cz}
\author{Karol Kampf,}\emailAdd{karol.kampf@mff.cuni.cz}
\author{David Podiv\'in}\emailAdd{david.podivin@mff.cuni.cz}
\affiliation{Institute for Particle and Nuclear Physics, Charles University, Prague, Czech Republic}

\abstract{
    We present a simple derivation of MHV amplitudes in massless spinor and scalar electrodynamics. Working with permutationally invariant amplitudes, we show that they are fully determined by their soft photon behavior and admit a simple factorized form in terms of soft factors and lower-point amplitudes. We prove these formulae using recursion relations. Finally, we consider possible extensions of these results by looking at supersymmetric theories, amplitudes beyond the MHV sector, gravity, and theories with charged particles of higher spins.
    }

\begin{document}

\maketitle

\section{Introduction}

The study of scattering amplitudes has seen an enormous amount of progress in the past several decades. The tools developed in this effort provide unprecedented computational power. This has led not only to an increased accuracy of theoretical predictions for experiments but also to completely new ways of understanding quantum field theories. The origins and most of the initial progress of this program were tied to Yang-Mills (YM) theory, examples being the seminal work by Parke and Taylor \cite{Parke1986} or the on-shell BCFW recursion relations \cite{Britto2005}. However, the applicability of these methods has been successively broadened to many other areas such as quantum electrodynamics (QED) \cite{Ozeren2005,Badger2010}, effective field theories \cite{Cheung2016,Cheung:2016drk,Bijnens:2019eze,Bartsch:2024amu}, gravity, supersymmetric theories, and many others. For recent review see e.g. \cite{Travaglini:2022uwo}.

We continue in a related effort to explore the structure of scattering amplitudes in various abelian gauge theories. Since amplitudes in the QED-like theories we investigate, unlike non-abelian theories like YM, do not admit a notion of color ordering, we are led to study permutationally invariant amplitudes directly. 
The goal of this article is to show that even fully permutationally symmetric MHV amplitudes exhibit a simple structure that is completely determined by their soft behavior. To accomplish this we employ BCFW recursion, manifesting permutational symmetry at every step of the calculation. This will directly yield a set of compact formulae for several processes in spinor QED and related theories. For each such process, we provide a validation of employed shifts in a dedicated Appendix.

\section{Quantum electrodynamics}\label{QED}

We start out by discussing scattering in renormalizable abelian gauge theories of massless charged particles with spin up to $s=1/2$.
Our initial focus will be on MHV amplitudes in spinor electrodynamics (QED) with two and four charged fermions. Although highly compact results are known for the two-fermion case \cite{Badger2010}, we will demonstrate that amplitudes involving four fermions exhibit the same level of simplicity. 
We then show that the general structure of these amplitudes carries over to scalar electrodynamics (sQED). All results are proven inductively using BCFW recursion \cite{Britto2005}.

\subsection{Spinor electrodynamics}

Amplitudes in spinor electrodynamics are known to be determined by just a single interaction vertex coupling two spin-$1/2$ fermions to a photon,
\begin{equation}\label{QED_vertex}
\vcenter{\includegraphics{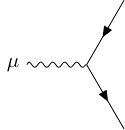}} \hspace{-12cm} = -i e\gamma^\mu.
\end{equation}
For amplitudes involving only a small number of particles this makes calculations tractable even in the conventional Feynman diagram formalism. However, we will show that the full simplicity of certain amplitudes can be exposed at arbitrary particle multiplicity when on-shell methods are used.
When computing amplitudes throughout this article we will consider all particles to be incoming. In diagrams we will also add arrows on lines corresponding to charged particles to indicate charge flow.

Let us first make a few general remarks about the structure of QED amplitudes involving both fermions and photons.
Tree-level diagrams in QED consist of fermion lines connected by photon propagators. Picking any external fermion line and following it through a diagram with an arbitrary number of photon lines gives the following expression
\begin{equation}
    \vcenter{\includegraphics{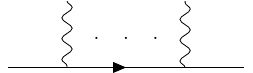} } \hspace{-10cm} = \Bar{v}(p_i) \Gamma u(p_j). 
\end{equation}
Here $\Gamma$ is a Dirac matrix and $p_i,p_j$ are momenta of the external fermions. Writing the relevant Feynman rule in spinor helicity form, the above diagram gives one of the following expressions \cite{Elvang2013} depending on the helicities of fermions $i$ and $j$,
\begin{equation}\label{zeroFerm}
\left[ i\right| \Gamma \left| j \right> \qquad \text{or} \qquad  \left< i\right| \Gamma \left| j \right] \qquad \text{or} \qquad \left[ i\right| \Gamma \left| j \right] = 0 \qquad \text{or} \qquad \left< i\right| \Gamma \left| j \right> = 0.
\end{equation}
Combining vertices \eqref{QED_vertex} and fermion propagators always produces an odd number of $\gamma$ matrices in $\Gamma$. Consequently, the last two expressions in \eqref{zeroFerm} identically vanish. This implies that in QED any fermion line must correspond to a fermion-antifermion pair with opposite helicity. 

With this in mind, there are four non-vanishing fundamental three-point amplitudes in spinor QED
\begin{equation}\label{qed3}
	\begin{matrix}
		A_3\left(f^-\Bar{f}^+\gamma^-\right) = e\sqrt{2}\frac{\left<13\right>^2}{\left<12\right>}, \quad A_3\left(f^+\Bar{f}^-\gamma^-\right) = e\sqrt{2} \frac{\left<23\right>^2}{\left<21\right>}, \\[0.2cm]
			 A_3\left(f^-\Bar{f}^+\gamma^+\right) =  e\sqrt{2} \frac{\left[23\right]^2}{\left[12\right]}, \quad A_3\left(f^+\Bar{f}^-\gamma^+\right) =  e\sqrt{2}\frac{\left[13\right]^2}{\left[21\right]},
	\end{matrix}
\end{equation}
where $\bar{f},f$ and $\gamma$ denote (anti-)fermions and photons. The superscript $\pm$ indicates helicity. These amplitudes will serve as input in the following calculations.

\subsubsection*{Two charged fermions}
Let us now consider the process $f \Bar{f} + N \gamma \to 0$. Original results for the corresponding amplitudes were obtained in \cite{Kleiss1986} using conventional Feynman diagram methods. For this process there is only one Feynman diagram topology and the amplitude is simply given by 
\begin{equation}\label{KSFD}
	iA = \hspace{-0.5cm}
\vcenter{\includegraphics{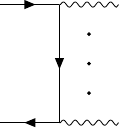}} \hspace{-12.3cm}
	+ \text{permutations of photon momenta}.
\end{equation} 
While Feynman diagrams produce a reasonably compact result for this particular amplitude, on-shell methods can be used to find even simpler expressions. This was demonstrated in \cite{Badger2010} by deriving a procedure to generate all helicity amplitudes for processes with two charged fermions. 
In particular, we quote here the following known two-fermion amplitudes \cite{Kleiss1986,Ozeren2005,Badger2010}
\begin{equation}
    	A\left(f^{h_1} \bar{f}^{h_2} \gamma^+ \ldots \gamma^+\right) = 	A\left(f^{h_1} \bar{f}^{h_2} \gamma^- \ldots \gamma^-\right)=0.
\end{equation}
as well as the MHV amplitude
\begin{equation}\label{KSS}
	A_n\left(f^- \bar{f}^+ \gamma^- \gamma^+ \ldots \gamma^+\right) = \frac{\spag{13}^2}{\spag{12}} \prod_{k=4}^{n} \frac{\spag{12}}{\spag{1k}\spag{2k}},
\end{equation}
which we will use for the subsequent analysis of amplitudes involving more than two fermions. In \eqref{KSS} the labels $1,2$ correspond to the momenta of the fermion-antifermion pair, $3$ to the negative ($-$) helicity photon and $4$ and higher to positive ($+$) helicity photons. Furthermore, we have set the electromagnetic coupling to $e \equiv 1 /\sqrt{2}$ for simplicity. It can always be uniquely restored by multiplying the result by $\left( \sqrt{2} e\right)^{n-2}$ if necessary.

\subsubsection*{Four charged fermions}
A natural question to ask is whether the simplicity of the MHV amplitude \eqref{KSS} can be extended to the scattering of four or more charged scalars.  
We therefore turn to the discussion of MHV amplitudes for the process $f \Bar{f} f \Bar{f} + N\gamma \to 0$ involving four charged fermions. There are two non-trivial helicity configurations that correspond to two fermion helicities $f^- \Bar{f}^+ f^+ \Bar{f}^-$ and $f^- \Bar{f}^+ f^- \Bar{f}^+$ and all photons with positive helicity. All other configurations can be obtained by parity or charge conjugation. This classification follows from \eqref{zeroFerm}.

To obtain results for an arbitrary number of photons we proceed as follows: first, we explicitly compute some lower point amplitudes. Based on their structure, we propose an ansatz for any multiplicity which we then prove by induction using recursion relations.

\paragraph*{Fermion helicity \texorpdfstring{\( (-++-) \)}{-++-}}

\begin{figure}[ht]
   	\centering
	\includegraphics[width=0.45\linewidth]{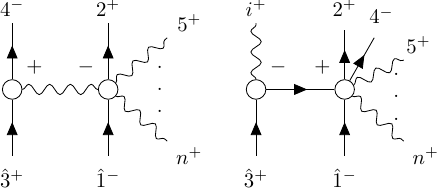}
	\caption{Factorization diagrams for process  $ f^-\bar{f}^+ f^+ \bar{f}^- + N \gamma ^+ \to 0$.  The only other amplitudes that appear on the factorization channels are the MHV amplitudes \eqref{KSS} involving two fermions.}
	\label{fig:diags1}
\end{figure}

To calculate the MHV process $f^-\bar{f}^+ f^+ \bar{f}^- + N \gamma^+ \to 0$, we choose a standard BCFW shift of the fermion momenta 
\begin{equation}\label{shf}
	\left|\hat{1}\right]= \left|1\right] + z \left| 3\right], \qquad 	\left|\hat{3}\right> = \left|3\right> - z \left| 1\right>.
\end{equation} We provide proof that this shift is valid in Appendix \ref{apd2}. The above shift produces two topologies of factorization diagrams shown in figure \ref{fig:diags1}.

Explicit calculations up to six points lead to
\begin{equation}
	A\left(f^-\Bar{f}^+f^+\Bar{f}^-\right) = \frac{\spag{14}^2}{\spag{12}\spag{34}},
\end{equation}
\begin{equation}
	A\left(f^-\Bar{f}^+f^+\Bar{f}^-\gamma^+\right)=\spag{14}^2 \left(\frac{1}{\spag{12}\spag{35}\spag{45}} + \frac{1}{\spag{15}\spag{25}\spag{34}}\right),
\end{equation}
\begin{multline}\label{4f6p1}
	A\left(f^-\Bar{f}^+f^+\Bar{f}^-\gamma^+ \gamma^+\right) = \spag{14}^2\left(\frac{\spag{12}}{\spag{34}\spag{15}\spag{16}\spag{25}\spag{26}}  +\frac{\spag{34}}{\spag{12}\spag{35}\spag{36}\spag{45}\spag{46}} \right.\\ + \left. \frac{1}{\spag{15}\spag{25}\spag{36}\spag{46}}+ \frac{1}{\spag{16}\spag{26}\spag{35}\spag{45}}\right).
\end{multline}
We provide a seven-point result in the Appendix \ref{7f} for completeness and better illustration of the general pattern upon which we now wish to expand.

Firstly, we see that photon momenta only appear in the denominator. Furthermore, in each term a given photon momentum only ever appears in spinor brackets together with a specific fermion-antifermion pair. E.g. in the first term of \eqref{4f6p1} photon $5$ is associated to the fermion pair $(1,2)$ via the brackets $\spag{15},\spag{25}$ and similar for photon $6$. In the second term, both photons $5$ and $6$ are associated with the fermion pair $(3,4)$. In the third term, photon $5$ appears in brackets only with fermions $(1,2)$, while photon $6$ is associated only to fermions $(3,4)$ and so on.

Following this pattern we propose an all-multiplicity expression as a sum over all possible ways in which photons can be associated to fermion pairs $(1,2)$ and $(3,4)$,
\begin{equation}\label{RES1}
	A\!\left( f^-\bar{f}^+ f^+ \bar{f}^-\gamma^+\ldots \gamma^+\right) = \spag{14}^2 \!\sum_{K\subseteq \gamma} \frac{\spag{12}^{|K|-1}\spag{34}^{|\gamma\setminus K|-1}}{\prod_{k\in K} \spag{1k}\!\spag{2k}\prod_{l\in \gamma\setminus K}\spag{3l}\!\spag{4l}}.
\end{equation}
In this formula the sum runs over all subsets $ K \subseteq \gamma$ of photon labels $\gamma=\lbrace 5\dots n\rbrace$. For notational convenience we define
\begin{equation}\label{T}
    T^K(a,b) = \frac{\spag{ab}^{|K|-1}}{\prod_{k \in K} \spag{ak}\!\spag{bk}},
\end{equation}
where $a,b$ are labels of a charged fermion and anti-fermion respectively and $K\subseteq \gamma$ is some subset of photon labels. Using this notation the expression \eqref{RES1} can be compactly written as
\begin{equation}\label{RES1T}
    A\!\left( f^-\bar{f}^+ f^+ \bar{f}^-\gamma^+\ldots \gamma^+\right) = \spag{14}^2 \!\sum_{K\subseteq \gamma} T^K(1,2)\,T^{\overline{K}}(3,4),
\end{equation}
where we denote by $\overline{K}=\gamma\setminus K$ the complementary set of $K$ in $\gamma$. 

Let us now proceed with the inductive proof of the above formula. 
The first diagram in Figure \ref{fig:diags1} can be evaluated using \eqref{KSS} and gives
\begin{align}\label{4f_diag_1}
	\begin{split}	\vcenter{\includegraphics[width=0.22\linewidth]{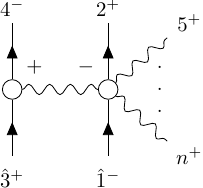}}\hspace{-10.7cm} = \spag{14}^2\frac{\spag{12}^{n-5}\spag{34}^{-1}}{\prod_{k=5}^n \spag{1k}\!\spag{2k}} = \spag{14}^2 T^{\gamma}(12)\, T^{\emptyset}(34),
	\end{split}
\end{align}
which exactly corresponds to the term in the sum \eqref{RES1T} where $K = \gamma $ and $ \overline{K} =\emptyset $.

The second set of diagrams in Figure \ref{fig:diags1} requires slightly more work. Summing all of them leads to
\begin{equation}\label{diags2}
	\sum_{i = 5}^n  \hspace{-0.5cm} \vcenter{\includegraphics[width=0.22\linewidth]{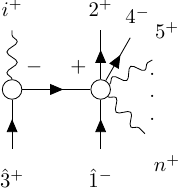}}\hspace{-11.2cm} =
    \spag{14}^2\!\sum_{i = 5}^n \sum_{K_i\subseteq \gamma_i} T^{K_i}(1,2) \frac{1}{\spag{i3}}T^{\overline{K}_{\!i}}(i,4).
\end{equation}
For each contribution $i$ we sum over subsets $K_i\subseteq \gamma_i$ of photon labels $\gamma_i = \{5,\ldots,\cancel{i},\ldots,n\}$ with the $i$-th label removed. As before, the notation $\overline{K}_{\hspace{-0.01cm}i} =\gamma_i\setminus K_i$ indicates the complement of the set $K_i$ in $\gamma_i$.
Performing the sum over $i$ one can show that
\begin{equation}\label{4fermSumSimpl}
   \spag{14}^2\!\sum_{i = 5}^n \sum_{K_i\subseteq \gamma_i} T^{K_i}(1,2) \frac{1}{\spag{i3}}T^{\overline{K}_{\!i}}(i,4) = \spag{14}^2\!\sum_{ K \subset \gamma}  T^K(1,2) \, T^{\overline{K}}(3,4)
\end{equation}
where on the right-hand side $K\subset \gamma$ runs over all proper subsets of photon labels.
Combining \eqref{4f_diag_1} and \eqref{4fermSumSimpl} we recover all the terms in \eqref{RES1T} which concludes the inductive step and proves the formula \eqref{RES1}.

\paragraph*{Fermion helicity \texorpdfstring{\( (-+-+) \)}{-+-+}}

\begin{figure}
	\centering
	\includegraphics[width=0.75\linewidth]{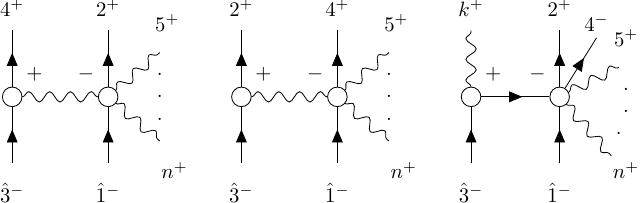}
	\caption{Factorization topologies of diagrams for  $ f^-\bar{f}^+ f^- \bar{f}^+  +N\gamma \to 0$. }
	\label{fig:diags2}
\end{figure}

Let us now turn our attention to the other fermion helicity configuration $ f^-\bar{f}^+ f^- \bar{f}^+ $. A key difference from the previous process is the presence of identical fermions. The corresponding statistics can be taken into account with a careful choice of relative signs between factorization diagrams. We once again employ the shift \eqref{shf}, and the relevant diagrams are depicted in Figure \ref{fig:diags2}.

Studying explicit expressions at low multiplicity, it turns out that the amplitudes of this process are given by a sum of two copies of the two-fermion result \eqref{RES1} with flipped fermion labels and a relative minus sign,
\begin{align}\label{RES2}
    \begin{split}
    A\!\left( f^-\bar{f}^+ f^- \bar{f}^+\gamma^+\ldots \gamma^+\right) = \spag{13}^2 \! \sum_{K\subseteq \gamma} \!\left\lbrace T^{K}(1,2)\, T^{\overline{K}}(3,4) - T^{K}(1,4)\, T^{\overline{K}}(3,2) \right\rbrace.
    \end{split}
\end{align}
where the sum over subsets $K\subseteq \gamma$ of photon labels is identical to the one in \eqref{RES1} and the factors $T^K(a,b)$ are defined in \eqref{T}.
The inductive proof for the expression \eqref{RES2} proceeds along the same lines as in the two-fermion case. We show that the employed shift \eqref{shf} is valid in Appendix \ref{apd2}.

\subsubsection*{BCFW and statistics}

Let us comment on a technical difficulty that arises when dealing with identical fermions in the context of BCFW recursion relations.
For bosons, the standard procedure of summing all BCFW factorization diagrams is sufficient to ensure the correct symmetry properties of the resulting amplitude. In particular, this is the case for applications to gluon scattering or supersymmetric theories with bosonic superfields commonly found in the literature. 
However, for our purposes, we have to ensure correct fermionic statistics by a careful choice of relative signs between factorization diagrams.

This can be achieved as follows. Firstly, it will be convenient to slightly change our notation for amplitudes. We split photons and (anti-)fermions according to their helicity and write the amplitude as
\begin{equation}\label{ampSplitLabel}
    A\big(\gamma^+| \gamma^- |  f^- |  \bar{f}^+|  f^+ | \bar{f}^- \big).
\end{equation}
The sets $\gamma^\pm,f^{\pm},\bar{f}^{\pm}$ contain the particle labels for each type, which we take to be in ascending order. To give an example, using the notation \eqref{ampSplitLabel} we can write the five-point amplitude from \eqref{RES2} as $A\!\left( f^- \bar{f}^+ f^- \bar{f}^+ \gamma^+\right) \equiv A\big(5^+ \vert \emptyset| 1^-  3^-| 2^+  4^+ |\emptyset|\emptyset \big)$. The empty set $\emptyset$ indicates the absence of negative helicity photons $\gamma^-$ and fermions $f^+\!,\bar{f}^-$ in the amplitude.

In the next step, we establish a fermion-antifermion pairing. We choose to pair fermions types $(f^-,\bar{f}^+)$ and $(f^+,\bar{f}^-)$ which we indicate as
\begin{equation}\label{ampPairings}
    A\big(\gamma^+| \gamma^- | \wick[sep = 0pt, offset = 1.5em]{ \c1 f^- |  \overline{ \c1 f}^+|  \c2 f^+ | \overline{\c2 f}^- }\big).
\end{equation}
By convention, we pair fermions in order $(f_i^-,\bar{f}_i^+)$ and $(f_j^+,\bar{f}_j^-)$. Since each set of particle labels is in ascending order, this pairing is unambiguous. 

For the five-point example above, we can therefore write \( A\big(5^+ \vert \emptyset|\wick{ \c1 1^- \c2 3^-| \c1 2^+ \c2 4^+ }|\emptyset|\emptyset \big)\), corresponding to the pairings $(1^-,2^+)$ and $(3^-,4^+)$.
Suppose that we want to compute this amplitude using a $\langle 13]$-shift,
\begin{align}\label{sign2}
	\begin{split}
		&\hspace{5cm} A\big(5^+ \vert \emptyset|\wick{ \c1 1^- \c2 3^-| \c1 2^+ \c2 4^+ }|\emptyset|\emptyset \big) = \\ &
    \vcenter{\includegraphics[width=0.8\linewidth]{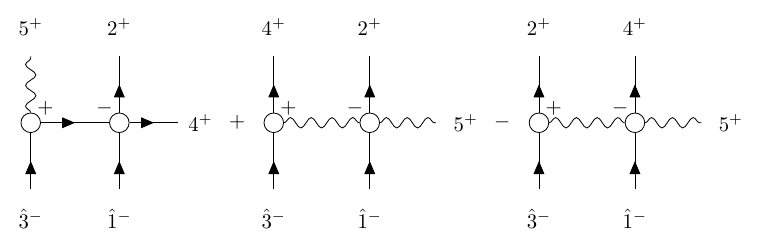}}\hspace{-3cm}
 	\end{split}\hspace{0.6cm}.
\end{align}
The factorization diagrams are built out of the following lower-point amplitudes
\begin{align}\label{sign3}
    \begin{split}
   &\text{First diagram:} \qquad A\big( 5^+ \vert \emptyset \vert \wick{ \c1 3^-  | \c1 P^+} \vert \emptyset \vert \emptyset  \big) A\big( \emptyset \vert \emptyset \vert \wick{\c1 1^- \c2 P^- \vert\c1 2^+ \c2 4^+  \vert \emptyset \vert \emptyset} \big) \to +1,\\
        &\text{Second diagram:} \hspace{-0.4cm}\qquad A \big( P^+ \vert \emptyset  \vert \wick{ \c1 3^- \vert \c1 4^+} \vert \emptyset \vert \emptyset\big)
        A \big( 5^+ \vert P^- \vert \wick{ \c1 1^- \vert \c1 2^+} \vert \emptyset \vert \emptyset \big) \hspace{0.42cm}\to +1,\\
        & \text{Third diagram:} \hspace{-0.2cm}\qquad A \big(P^+ \vert \emptyset  \vert \wick{ \c1 3^- \vert \c1 2^+} \vert \emptyset \vert \emptyset\big)
        A \big( 5^+ \vert P^- \vert \wick{ \c1 1^- \vert \c1 4^+} \vert \emptyset \vert \emptyset \big) \hspace{0.42cm} \to -1,
    \end{split}
\end{align}
where we adhere to the same conventions regarding fermion pairings for the sub-amplitudes in each diagram. We can then determine the sign of a given diagram by comparing its pair assignments to those of the full amplitude we want to compute.

To illustrate this, let us consider the diagrams \eqref{sign2} in turn. From \eqref{sign3} we see that the first diagram corresponds to pairings $(3^-,P^+),(P^-,4^+)$ and $(1^-,2^+)$. To obtain pairings only of external particle labels we can contract pairings involving the same intermediate state, i.e. in our case we can eliminate $P^\pm$ and set $(3^-,P^+),(P^-,4^+)=(3^-,4^+)$. Thus we end up with pairs $(1^-,2^+),(3^-,4^+)$ for the first diagram, which are the same as for the original amplitude. By our convention, it therefore carries a relative $+$ in \eqref{sign2}. We can treat diagram two in \eqref{sign3} using exactly the same logic as for diagram one.
For the third diagram we immediately read off the pairings $(1^-,4^+),(3^-,2^+)$ from \eqref{sign3}. In this case no contractions of intermediate states need to be taken into account as the exchanged particle is a photon. Now, starting from the pairings of diagram three we have to perform exactly one transposition of labels ($2^+ \!\leftrightarrow 4^+$) to arrive back at the pairings of the original amplitude. Each such transposition introduces a factor $(-1)$ to the given diagram, which explains the relative minus sign of the third diagram in \eqref{sign2}.

While the above example should provide enough detail to consistently implement BCFW recursion with identical fermions at any multiplicity, let us briefly comment on the general case. Let us consider a generic amplitude \eqref{ampPairings} with fermion pairings $(f_i^-,\bar{f}_i^+)$ and $(f_j^+,\bar{f}_j^-)$ in standard order. Computing the amplitude from BCFW, a given factorization diagram will have pair assignments which can be always be written as $( f^-_i , \bar{f}^+_{\sigma(i)})$ and $(f^+_j , \bar{f}^-_{\pi(j)} )$, where $\sigma,\pi$ are some permutations of the pairing labels $i,j$. The relative sign of the diagram is then given by the sign of the two permutations \((-1)^{\sigma + \pi}\).

\subsubsection*{Soft factor form for MHV amplitudes}

The previously derived formulae can be even further simplified by taking advantage of their known soft behavior. In QED after rescaling momentum of a photon to $k \to \varepsilon k$ one finds the soft limit
\begin{equation}\label{photSftThm}
A\left(f_1 \ldots \bar{f}_m \gamma_{m+1} \ldots \gamma_k \ldots \gamma_n\right) \overset{ \varepsilon \to 0}{\to} \frac{1}{\varepsilon} S_k^{(m)}  A\left(f_1 \ldots \bar{f}_m \gamma_{m+1} \ldots \cancel{\gamma_k} \ldots \gamma_n\right) + O(\varepsilon^0)
\end{equation}
Explicitly, for an amplitude with an even number $m$ of charged particles the eikonal/soft factor $S^{(m)}_k$ obtained by taking photon $k$ soft is given by
\begin{equation}\label{eik}
	S_k^{(m)} = \frac{1}{\sqrt{2}} \!\sum_{i=1}^{m}q_l \frac{\epsilon\left(k , +\right) \!\cdot p_i}{k \cdot p_i} = \sum_{\substack{j=1\\\text{odd}}}^{m-1}\frac{\spag{j\,j{+}1}}{\spag{jk}\spag{j{+}1\,k}}
\end{equation}
Now the four-fermion amplitudes \eqref{RES1} and \eqref{RES2} can be rearranged to take the form
\begin{align}\label{SOFT12}
    \begin{split}
        A\!\left( f^-\bar{f}^+ f^+ \bar{f}^-\gamma^+\ldots \gamma^+\right) &=\frac{\spag{14}^2}{\spag{12}\spag{34}}\prod_{k=5}^{n}\!\left(\frac{\spag{12}}{\spag{1k}\spag{2k}}+\frac{\spag{34}}{\spag{3k}\spag{4k}}\right),\\
        A\!\left( f^-\bar{f}^+ f^-\bar{f}^+\gamma^+\ldots \gamma^+\right) &=\frac{\spag{13}^3\spag{24}}{\spag{12}\spag{23}\spag	{34}\spag{41}}\prod_{k=5}^{n}\!\left(\frac{\spag{12}}{\spag{1k}\spag{2k}}+\frac{\spag{34}}{\spag{3k}\spag{4k}}\right).
    \end{split}
\end{align}
Both exhibit a simple factorized structure in terms of the basic four-point amplitude involving fermions of the given helicity multiplied with an eikonal factor \eqref{eik} for each photon,
\begin{equation}\label{FullFermion}
		A\!\left( f^{h_1}\bar{f}^{h_2} f^{h_3}\bar{f}^{h_4}\gamma^+\ldots \gamma^+\right) =	A\!\left( f^{h_1}\bar{f}^{h_2} f^{h_3}\bar{f}^{h_4}\right) \!\prod_{k \in \text{photons}}\!\!\! S^{(4)}_k,
\end{equation}
Let us remark on a subtlety regarding momentum conservation for the four-point factor on the right-hand side. Since the fermion momenta do not obey momentum conservation by themselves when $n\ge 5$ the validity of \eqref{FullFermion} hinges on choosing a preferred functional form of the four-fermion amplitude. The specific four-point functions shown in \eqref{SOFT12} provide such a preferred choice.

As we will show, the soft factor structure \eqref{FullFermion} turns out to apply to a large class of MHV amplitudes in a wide range of theories. In fact, the two-fermion MHV amplitude \eqref{KSS} encountered earlier can also be recognized to take the factorized form
\begin{equation}\label{FullFermion2}
		A\!\left( f^{-}\bar{f}^{+} \gamma^-\gamma^+\ldots \gamma^+\right) =	A\!\left( f^{-}\bar{f}^{+} \gamma^-\right) \!\prod_{k \in \text{photons}}\!\!\! S^{(2)}_k,
\end{equation}
where the three-point factor is given by \eqref{qed3} and $S^{(2)}_k$ is the soft factor for two charged fermions.
While the soft factor form \eqref{FullFermion2} for the particular case of electron-positron annihilation has been known in the literature for some time, we have shown here that this structure extends to MHV processes with four fermions. Similar results for four fermions have been derived for color-ordered MHV amplitudes in QCD by projecting components from $\mathcal{N} {=} 4$ superamplitudes \cite{Dixon2010}. The form \eqref{SOFT12} can also be found by removing color factors from the QCD results \cite{Mangano2005}, but the results given there are only valid for different fermion flavors with identical charges.

\paragraph*{Different charges}

Assuming that the fermion pairs have different charges, the previous procedure is very simple to repeat and recover the formula \eqref{FullFermion} with only a simple change in the eikonal factors to
\begin{equation}\label{FullSoft}
     \left( q \frac{\spag{12}}{\spag{1k}\spag{2k}}+ Q\frac{\spag{34}}{\spag{3k}\spag{4k}}\right),
\end{equation}
where $q,Q$ are the respective charges of the particles in units of electron charge.

\subsection{Scalar electrodynamics}

Let us now turn to the electrodynamics of massless charged scalars. Our aim is again to derive compact formulae for MHV amplitudes involving eikonal factors analogous to those found in spinor QED \eqref{FullFermion}. 

Following the example of spinor QED, we want to study amplitudes in the most general renormalizable theory with dimensionless couplings, now involving photons and a charged scalar. This leads us to consider a theory with two coupling constants described by the Lagrangian
\begin{equation}\label{scalar_lag}
	\mathcal{L} = -\frac{1}{4} F^{\mu \nu}F_{\mu \nu} + \left(D_\mu \phi\right)^\dagger  \left(D^\mu \phi\right) - \frac{\lambda}{4} \left(\phi^\dagger \phi\right)^2,
\end{equation}with the standard covariant derivative $ D_\mu = \partial_\mu - i e A_\mu $. From the Lagrangian we can read off three interaction vertices

\begin{equation}\label{scalar_FR}
    \vcenter{\includegraphics{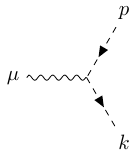}} \hspace{-12.3cm} = - ie(p-k)_\mu,\vcenter{\includegraphics[width = 0.15\linewidth]{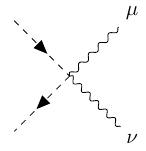}} \hspace{-12.3cm} = 2 ie^2 \eta_{\mu\nu},\vcenter{\includegraphics[width = 0.15\linewidth]{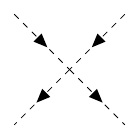}} = \hspace{-13cm} = -i\lambda.
\end{equation}
The two relevant three point amplitudes are
\begin{equation}\label{3scalar}
	A_3\left( \phi \phi^* \gamma^-\right) = \frac{\spag{13}\spag{23}}{\spag{12}}, \qquad A_3\left( \phi \phi^* \gamma^+\right) = \frac{\spsq{13}\spsq{23}}{\spsq{12}}.
\end{equation}
Due to the existence of the four scalar contact interaction \eqref{scalar_FR}, this theory cannot be reconstructed fully from these three-point amplitudes alone as the contact interaction does not appear in any factorization channel at four points. We thus have to take four-point amplitudes as an additional input. This issue only arises for amplitudes involving four or more scalars as the couplings of all other interaction terms are fixed by gauge invariance. 

\subsubsection*{Two charged scalars}

Looking at the previously derived soft factor forms \eqref{SOFT12} and \eqref{FullFermion2} of the MHV amplitudes in spinor QED, a natural guess for the two-scalar amplitude is
\begin{equation}\label{ResS1}
	A\!\left(\phi \phi^*\gamma^- \gamma^+ \ldots \gamma^+ \right) = A\!\left( \phi \phi^* \gamma^- \right) \prod_{k=4}^n S^{(2)}_k = \frac{\spag{13}\spag{23}}{\spag{12}} \prod_{k=1}^n \frac{\spag{12}}{\spag{1k}\spag{2k}}.
\end{equation}

A crucial ingredient in the proof of the above expression will be that amplitudes with all equal helicity photons vanish 
\begin{equation}\label{S0}
		A\!\left(\phi \phi^*  \gamma^+ \ldots \gamma^+ \right) = A\!\left(\phi \phi^*  \gamma^- \ldots \gamma^- \right) = 0,
\end{equation}
which will ensure the absence of three- and higher particle multiplicity channels in the MHV amplitude.
Performing the shift 
\begin{equation}\label{scalar_shift_2}
	\left|\hat{1}\right> = 	\left|1\right> - z 	\left|2\right>, \qquad \left|\hat{2}\right] = 	\left|2\right] + z 	\left|1\right],
\end{equation}
and using \eqref{S0} there is only one diagram topology to evaluate
\begin{align}
	\begin{split}\label{ID2}
			A\left(\phi \phi^* \gamma^- \gamma^+ \ldots \gamma^+\right) = \sum_{k=4}^n \begin{matrix} \vspace{-0.2cm} \includegraphics[width=3.00cm]{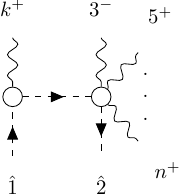} \end{matrix} \hspace{-0.1cm} &= \sum_{k=4}^n  \frac{\spag{21}\spag{k3}\spag{23}}{\spag{k2}^2\spag{k1}}\prod_{l\neq k} \frac{\spag{k2}}{\spag{kl}\spag{2l}}\\  &=  \frac{\spag{13}\spag{23}}{\spag{12}} \prod_{k=1}^n \frac{\spag{12}}{\spag{1k}\spag{2k}}.
    \end{split}
\end{align}
Here, the last equality follows from the generalized Schouten identity \eqref{genschout}. This concludes the inductive proof up to the verification of the shift, which can be found in Appendix \ref{apd2}.

Let us quickly comment on related results obtained from Feynman diagrams. In fact, for two charged scalars a closed formula for the amplitudes is known \cite{Badger2008} for any helicity configuration. However, the expressions derived here for MHV amplitudes are more compact and better suited for recursive computations of more involved amplitudes later on.

\subsubsection*{Four charged scalars}
For four charged scalars, we follow the previous result \eqref{SOFT12} and make the simple ansatz
\begin{equation}\label{soft_4_scalar}
    A\!\left(\phi \phi^* \phi \phi^*  \gamma^+ \ldots \gamma^+ \right) = A\!\left( \phi \phi^* \phi \phi^*\right) \prod_{k=5}^n S^{(4)}_k
\end{equation}
Where the soft factors $S^{(4)}_k$ are as in \eqref{eik} and the four-scalar amplitude is given by
\begin{equation}\label{4scal}
		A\!\left(\phi \phi^* \phi \phi^* \right) = \frac{\spag{13}^2\spag{24}^2}{\spag{12}\spag{14}\spag{32}\spag{34}} + C.
\end{equation}
 The dimensionless constant $ C $ is related to the Lagrangian parameters in \eqref{scalar_lag} by $ C = 2e^2-\lambda \equiv 1-\lambda $, where we recall that we set $ e= 1/\sqrt{2} $ for convenience. Although the constant $\lambda$ is arbitrary in general, some additional symmetries might restrict it. One such possibility is supersymmetry, which enforces $\lambda = 2e^2$ or $C=0$, a case we will consider in more detail later.

\begin{figure}
	\centering
	\includegraphics[width=0.75\linewidth]{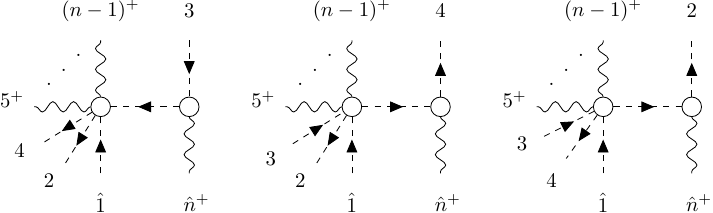}
	\caption{Factorization diagrams for $ \phi \phi^* \phi \phi^* + N\gamma^+ \to 0 $.}
	\label{fig:ks4}
\end{figure}

Moving on to the inductive proof of the four scalar amplitude \eqref{soft_4_scalar}, the shift we now employ is
\begin{equation}\label{SShift}
		\left|\hat{1}\right]= \left|1\right] + z \left| n\right], \qquad 	\left|\hat{n}\right> = \left|n\right>- z \left| 1\right>.
\end{equation}
The choice of the photon label $ n $ is arbitrary due to the symmetry of the underlying amplitude, but it is convenient for notational purposes. Only three factorization diagrams are nonzero for this process and are shown in Figure \ref{fig:ks4}. All other topologies vanish by three-point kinematics or \eqref{S0}.

Evaluating each non-zero diagram explicitly results in an expression involving the same four-scalar amplitude $A_{n-1}(\phi \phi^* \phi \phi^*  \gamma^+ \ldots \cancel{\gamma^+})$ with the $ n $-th photon removed multiplied by a simple kinematic factor specific to the diagram. Summing the contributions from all diagrams gives
\begin{equation}
	A_n = \underbrace{A_4\prod_{ k =5}^{n-1}\Bigg(\frac{\spag{12}}{\spag{1k}\spag{2k}} + \frac{\spag{34}}{\spag{3k}\spag{4k}}\Bigg)}_{A_{n-1}}
    \underbrace{\Bigg( \overbrace{-\frac{\spag{13}}{\spag{1n}\spag{3n}}}^{\text{Diag. 1}}+\overbrace{\frac{\spag{14}}{\spag{1n}\spag{4n}}}^{\text{Diag. 2}} +\overbrace{\frac{\spag{12}}{\spag{1n}\spag{2n}}}^{\text{Diag. 3}} \Bigg)}_{S_n^{(4)}},
\end{equation}
where the terms neatly combine into the soft factor for $k=n$, validating the formula \eqref{soft_4_scalar}.

\subsubsection*{Other scalar amplitudes}
Let us briefly comment on some other amplitudes involving scalars that exhibit a simple soft factor structure.
Considering for instance the case of two flavors of complex scalars $\phi$ and $\varphi$, the formula \eqref{soft_4_scalar} still holds with the four-point amplitude replaced by
\begin{equation}
    A\!\left( \phi \phi^* \varphi \varphi^*\right) = -\frac{\spag{13}\spag{24}}{\spag{12}\spag{34}} + C.
\end{equation}
Here, the constant is again $C = 1- \lambda$ with the scalar coupling $\lambda$ coming from the potential interaction $ -\lambda \: \phi^\dagger \phi \: \varphi^\dagger \varphi $. If in addition the different flavors of scalars have different charges, only the soft factors have to be replaced by \eqref{FullSoft} as before.

The generic form of the amplitude \eqref{soft_4_scalar} even holds for a theory involving both fermions and scalars. The MHV amplitude $A\left(f^- \bar{f}^+ \phi \phi^* \gamma^+ \ldots \gamma^+\right)$ can be written in soft factor form where the relevant four-point process is simply
\begin{equation}\label{mix}
    A\!\left( f^- \bar{f}^+ \phi \phi^*\right) = \frac{\spag{13}\spag{14}}{\spag{12}\spag{34}}.
\end{equation}
Quite naturally, the comments about different charges still hold and modify only the soft factors.

\subsection{\texorpdfstring{$\mathcal{N} = 2$}{N = 2} supersymmetry}

Comparing the previous formulae with fermions \eqref{FullFermion} and scalars \eqref{soft_4_scalar} we see that they only differ by the four-point amplitudes. The fact that their structure is so similar can be seen by introducing supersymmetry, where both amplitudes can be combined into a single superamplitude. Concretely we want to study $\mathcal{N}=2$ super-QED with matter supermultiplets $\Phi, \bar{\Phi}$  combining scalars and fermions. The theory also contains massless vectors $V^{\pm}$ but does not allow for self-interactions. As we will see, this ensures that superamplitudes involving vectors exhibit a QED-like structure resembling previously derived results.

Using super-BCFW shifts \cite{Elvang2011} we find a familiar form of superamplitudes in the MHV sector. For two charged superfields we obtain
\begin{equation}\label{N2_C2}
	\mathcal{A}\!\left(\Phi \overline{\Phi} V^- V^+ \ldots V^+\right)= \frac{\gdf{4}}{\spag{12}}\prod_{k=4}^nS_k^{(2)}
\end{equation}
whereas the result for four charged superfields reads
\begin{equation}\label{N2_C4}
	\mathcal{A}\!\left(\Phi \overline{\Phi} \Phi \overline{\Phi} V^+ \ldots V^+\right)= \frac{\gdf{4}\spag{13}\spag{24}}{\spag{12}\spag{23}\spag{34}\spag{41}}\prod_{k=5}^n S_k^{(4)},
\end{equation}
both of which clearly take the known soft factor form. The only new feature of these superamplitudes are the Grassmann delta functions which conserve supermomentum.

Now we can extract the component amplitudes with up to four charged scalars and spinors. Firstly we find that the result with two charged fields \eqref{N2_C2} matches the non-supersymmetric results \eqref{KSS} and \eqref{ResS1}. Secondly, from the result with four charged scalars, we see that supersymmetry has enforced the constant term $C=0$ by relating the couplings $\lambda = 2e^2$ in the Lagrangian \eqref{scalar_lag}. From an on-shell perspective this specific choice of $C$ is required to ensure the absence of poles at infinity and therefore the consistency of the recursive calculation using BCFW.

Finally, we find a discrepancy when comparing the $\mathcal{N} = 2$ result with 4 charged superfields \eqref{N2_C4} and the four fermion amplitudes in \eqref{FullFermion}. This last difference is due to an additional scalar in the supersymmetric theory which contributes to the amplitude through a Yukawa-type interaction. This difference can be explicitly found in the four fermion amplitude
\begin{equation}
    A^{\mathcal{N} = 2} \left( f^- \bar{f}^+  f^+ \bar{f}^-\right) = \bigg(\underbrace{\frac{\spag{14}^2}{\spag{12}\spag{34}}}_{\text{photon}} + \underbrace{\frac{\spag{14}}{\spag{23}}}_{\text{scalar}} \bigg) \prod_{k=5}^n S_k^{(4)}
\end{equation}
However, thanks to the factorized soft factor form it is straightforward to project out from the four-point factor the piece corresponding to photon exchange and recover the desired non-supersymmetric QED result \eqref{SOFT12}.

\section{Electrodynamics of spins \texorpdfstring{$s \geq 1$}{s>1}}\label{s_1}

After studying charged matter with spins zero and one-half we would like to extend our discussion to particles with spin one or greater. The main obstacle we encounter when considering massless vectors is the Weinberg-Witten theorem \cite{Weinberg1980}, which states that massless particles with spins greater than one-half cannot carry conserved charges and thus cannot consistently couple to photons. 
In the following, we will first discuss several extensions of our previous results that avoid this theorem. We then illustrate how the Weinberg-Witten theorem appears from an on-shell perspective, where naive recursive calculations for amplitudes involving particles of spin $s>1$ lead to various inconsistencies.

\subsection{Spin one, Yang-Mills theory and supersymmetry}
 
One way to avoid the Weinberg-Witten theorem would be to consider massive charged vectors instead of massless ones. However, the minimal coupling of a massive spin one particle to the electromagnetic field is not renormalizable \cite{Horejsi2022}. This issue can be addressed by incorporating both vector and photon into a spontaneously broken Yang-Mills theory as e.g. in the electroweak sector of the standard model. Thus, having a consistent theory of massive charged spin one particles necessarily entails the presence of non-abelian gauge redundancy. On the other hand, non-abelian gauge theory by itself already bypasses the Weinberg-Witten theorem without the need to introduce masses by introducing non-abelian charges.

We therefore slightly broaden the scope of this article and, instead of insisting on purely abelian interactions, consider for simplicity a $SU(2)$ gauge theory of massless vector bosons. The shift to pure Yang-Mills theory also lends itself nicely to a discussion of supersymmetric amplitudes.

In fact, we will begin by considering the case of maximal supersymmetry, $\mathcal{N}{=}\,4$ super-Yang-Mills. The three superfields we encounter correspond to the $SU(2)$ generators $T^\pm, T^3$ and we shall denote them $W,\overline{W},V$. To make a connection to the previous results, let us call $W,\overline{W}$ charged fields and $V$ a "photon". 

We first compute the amplitude for two charged vector superfields $W,\overline{W}$ and an arbitrary number of neutral photons $V$. Using super-BCFW \cite{Elvang2011} shifts we find the familiar soft factor form previously encountered for fermions \eqref{KSS} and scalars \eqref{ResS1}, that is
\begin{equation}\label{SUSYKS2}
	\mathcal{A}^{\text{MHV}}\!\left(W \overline{W} V\ldots V\right) = \frac{\gdf{8}}{\spag{12}^2}\prod_{ k =3}^n S_k^{(2)}.
\end{equation}
Here, the Grassmann delta function $\gdf{2\mathcal{N}}$ enforces supermomentum conservation.

For four charged vector superfields, we arrive in a similar fashion at a formula analogous to \eqref{FullFermion}, \eqref{soft_4_scalar} which reads
\begin{equation}\label{SUSYKS4}
	\mathcal{A}^{\text{MHV}}\!\left(W \overline{W} W \overline{W} V\ldots V\right) = \frac{\gdf{8}}{\spag{12}\spag{14}\spag{32}\spag{34}} \prod_{ k =5}^n S_k^{(4)}.
\end{equation} 
By extracting the relevant component amplitudes from \eqref{SUSYKS2} and \eqref{SUSYKS4}, it is possible to rederive the results for fermions \eqref{SOFT12} and the two-scalar amplitude \eqref{ResS1}. However, projecting out the four-scalar amplitude from \eqref{SUSYKS4} we find 
\begin{equation}
    A\left(\phi \phi^* \phi \phi^* \gamma^+ \ldots \gamma^+\right) = \frac{\spag{13}^2\spag{24}^2}{\spag{12}\spag{14}\spag{32}\spag{34}} \prod_{k=5}^{n} S_k^{(4)}
\end{equation}
Comparing this to the previously derived result \eqref{soft_4_scalar}, we see that supersymmetry has once again chosen the constant term $C=0$ or equivalently $\lambda = 2e^2$ in the Lagrangian \eqref{scalar_lag}. 

Remarkably, the presence of maximal supersymmetry ensures that the soft factor form of \eqref{SUSYKS2} and \eqref{SUSYKS4} extends to amplitudes with any number of pairs of charged superfields,
\begin{equation}\label{SUSYKSN}
    \mathcal{A}^{\text{MHV}}_n \big(\underbrace{W \overline{W} \ldots W \overline{W}}_{2m} \hspace{1mm} V \ldots V \big) =  \mathcal{A}^{\text{MHV}}_{2m} \big(W \overline{W} \ldots W \overline{W} \big) \prod_{k=2m+1}^n S^{(2m)}_k.
\end{equation}
where $m$ denotes the number of ($W$,$\overline{W}$)-pairs. For $m=1,2$ the general formula \eqref{SUSYKSN} reduces to \eqref{SUSYKS2} and \eqref{SUSYKS4} respectively.

The amplitude involving $2m$ charged superfields $W,\overline{W}$ can be obtained by using the standard color-decomposition formula
\begin{equation}\label{full_W}
    \mathcal{A}^{\text{MHV}}_{2m} \!\left(W \overline{W} \ldots W \overline{W} \right) =\!\!\!\!\sum_{\pi \in S_{2m}/Z_{2m}}^{} \!\!\!\!\text{Tr}\!\left[T^{\pi(a_1)}\ldots T^{\pi(a_{2m})}\right] \,\mathcal{PT}\!\left[\pi (1)\ldots\pi (2m) \right].
\end{equation}
where the generators are $T^{\pi(a_i)}{\,=\,}T^{\pm}$ depending on whether the corresponding field is $W,\overline{W}$ and $\mathcal{PT}$ is the Parke-Taylor factor including a Grassmann delta function
\begin{equation}
    \mathcal{PT}\left(1\ldots 2m \right) = \frac{\gdf{8}}{\spag{12}\spag{34}\ldots \spag{2k{-}1 \; 2m}\spag{2m  \; 1}}.
\end{equation}
Let us note that the expression \eqref{full_W} has no equivalent in pure QED of the previous Section since the helicity structure of those theories does not allow them to exist.

\subsection{Higher charged spins}

Let us now try to go beyond spin one and construct amplitudes for particles of higher spin. We will employ the same on-shell techniques used previously to recursively build up amplitudes for particles with spin $s>1$. Due to the Weinberg-Witten theorem we expect this approach to eventually run into obstructions. These will appear as inconsistencies in the amplitudes derived through the recursion. We also discuss ways to amend some of these issues in certain cases. In our construction, we assume minimal electromagnetic coupling such that amplitudes depend only on a single dimensionless coupling constant. 

Three-point amplitudes involving particles of arbitrary helicity $h$ and photons are fixed from little group scaling. We then use recursion relations to formally compute amplitudes for any number of external legs. For helicities $|h|> 1$ we will find that the amplitudes computed in this way are inconsistent.

We can now proceed to compute $n$-point amplitudes using BCFW recursion.
Let us start with amplitudes for two charged particles with helicity $h$. We find
\begin{equation}\label{2hAmp}
    A\!\left( 1^h \bar{2}^{-h} \gamma^- \gamma^+ \ldots \gamma^+\right) = \frac{\spag{13}\spag{23}}{\spag{12}} \left(\!\frac{\spag{23}}{\spag{13}}\!\right)^{\!2h} \prod_{k=4}^n S_k^{(2)}.
\end{equation}
As anticipated, the amplitude is unphysical for $|h| > 1$ as it develops higher-order poles. Thus the Weinberg-Witten theorem appears as violation of locality in the higher spin amplitudes. Interestingly, however, the generic soft factor form still persists even in the unphysical case.  

Continuing with four charged particles we similarly obtain
\begin{equation}
    A\left(1^h \bar{2}^{-h} 3^s \bar{4}^{-s} \gamma^+ \ldots \gamma^+\right) = \frac{\spag{13}\spag{24}}{\spag{12}\spag{34}} \left(\! \frac{\spag{24}}{\spag{14}}\!\right)^{\!2h}\! \left(\! \frac{\spag{14}}{\spag{13}}\!\right)^{\!2s}\prod_{k=5}^n S_k^{(4)},
\end{equation}
and, assuming identically charged particles, a symmetrized form
\begin{equation}\label{4hAmp}
  A\left(1^h \bar{2}^{-h} 3^h \bar{4}^{-h} \gamma^+ \ldots \gamma^+\right) = \frac{\spag{13}^2\spag{24}^2}{\spag{12}\spag{14}\spag{32}\spag{34}} \left( \!\frac{\spag{24}}{\spag{13}}\!\right)^{\!2h} \prod_{k=5}^n S_k^{(4)}.
\end{equation}
Again, for $|h|,|s| > 1$ these formulae are not local.

\begin{figure}[ht]
    \centering
    \includegraphics[width=0.7\linewidth]{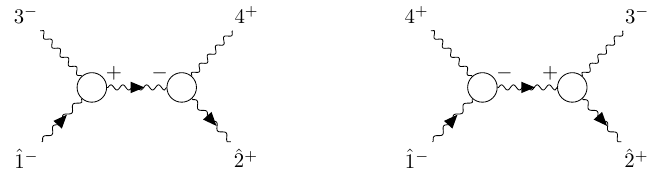}
    \caption{Factorization diagrams for Compton scattering in vector QED. The second diagram on the right only appears for charged particles with $h=1$.}
    \label{fig:spin1}
\end{figure}

In the present context let us briefly return to the case $h=1$, as the amplitudes \eqref{2hAmp},\eqref{4hAmp} still exhibit an inconsistent pole structure. This seems to contradict the consistent results we found for charged vectors in \eqref{SUSYKS2},\eqref{SUSYKS4}. 
A more careful analysis shows that for vectors in particular additional factorization diagrams appear in the recursion which were assumed not to be present in the computation of \eqref{2hAmp},\eqref{4hAmp}. These additional contributions cure the unphysical pole structure. As an example, we depict a consistent set of factorization diagrams for the case of Compton scattering in Figure \ref{fig:spin1}.

Let us now show that higher spin amplitudes cannot be cured in a similar way, at least not without introducing additional particles. For concreteness, we shall again focus on four-point Compton scattering of two charged particles with spins $h$ and two photons.

\begin{figure}[ht]
   \centering
   \includegraphics[width=0.7\linewidth]{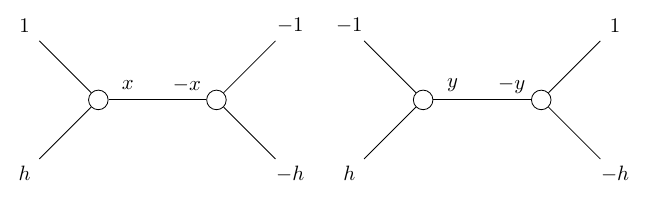}
   \caption{Two possible factorization diagrams for Compton scattering of particles with helicity $h$ and photons with helicities $\pm 1$. The internal labels $x,y$ are helicities of the propagating particles. We have assumed that charged particles are shifted.}
   \label{fig:h}
\end{figure}

We suppose that there are two factorization diagrams contributing to this process with helicities $x,y$ in intermediate states, as shown in Figure \ref{fig:h}. Now we know that the mass dimensions of any three-point amplitude is $\left[A_3\right]=1$. For a general three-point amplitude with helicities $h_1,h_2,h_3$ its mass dimension is related to the helicities via \cite{Elvang2013}
\begin{equation}
    1 = \left[ A_3\right] = \left[e\right] + \left| h_1 +h_2 +h_3\right|,
\end{equation}
where $e$ is the electromagnetic coupling for which $\left[e\right]=0$. Therefore the equality reduces to $\left| h_1 + h_2 + h_3\right| = 1$.
Applied to our example of Compton scattering we can derive four relations for the helicities $h$ and $x,y$ involved, one for each three-point amplitude appearing in the factorization diagrams in Figure \ref{fig:h},
\begin{equation}\label{x}
    h + x + 1 = 1, \quad  - x -h - 1 = -1,  \quad  h + y -1 = 1,  \quad  -h - y + 1 = -1.
\end{equation}
This set of equations has a unique solution $x = -h$ and $ y = 2 - h$. A special case is $h = 1$ as then $x = -y = -1$, which corresponds to Yang-Mills theory as discussed earlier. It is the only case where $|h|=|x|=|y|=1$ and additional particles with different spins do not have to be added to the theory.

Conversely, for $h>1$ new particles would have to be introduced as intermediate states in factorization diagrams which may cure the non-local pole structure for the given Compton process. However, finding a putatively consistent theory requires then to also consider amplitudes involving the newly added particles as external states and ensure their consistency. This in turn might necessitate the addition of even more particles and so on. Since this procedure is rather open-ended we refrain from pursuing it any further here.

\section{Extensions and generalizations}\label{ext}

\subsection{Soft factors beyond MHV}

So far we have studied MHV amplitudes in theories with charged matter of spins up to one and found that they all exhibit a simple structure involving products of soft factors and lower-point amplitudes. Afterwards, we have argued that this structure can be formally extended to charged particles of any spin. Naturally, one might ask how much of this simplicity extends to amplitudes beyond the MHV sector.

Firstly, let us clarify our terminology. An $n$-point amplitude is MHV if it has total helicity $n-4$.  However, QED processes with six or more charged particles already have their total helicity $n-6$ at most\footnote{Technically the amplitudes with higher total helicity exist in spinor QED but they identically vanish.}. This observation follows from our discussion around \eqref{zeroFerm} in spinor QED and is trivial for scalar QED. Thus, amplitudes with six or more charged particles classify as NMHV at the least.

Let us therefore look at base processes in QED-like theories involving six, eight, or even more fermions or scalars. We then start attaching eikonal factors to the amplitude of the base process and check if we obtain sensible results for the corresponding process with additional photons.
We shall proceed in a way that guarantees proper soft behavior. Naively we may expect that an amplitude involving photons is fully given by eikonal factors times lower point amplitudes summed over all photons that can go soft. Symbolically this reads $ A_n = \sum_k S_k A_k$, where $A_k=A\big(\cancel{\gamma^k}\big)$ denotes the $(n{-}1)$-point amplitude with photon $k$ removed. However, the above expression does not have the correct low-energy behavior as the soft theorems of the lower-point amplitudes $A_k$ themselves need to be taken into account. This can be accomplished by subtracting terms of the form $S_k S_l A_{kl}$, where $A_{kl}=A\big(\cancel{\gamma^k} \cancel{\gamma^l}\big)$ is now an $(n{-}2)$-point amplitude with two photons $k,l$ removed compared to the process described by $A_n$. We proceed in this manner until there are no photons left for us to remove. Consistently carrying out this procedure leads to the following function
\begin{equation}\label{soft_guess}
     A_n = \sum_k S_k A_k - \sum_{k < l} S_k S_l A_{kl} + \ldots + \prod_k S_k A_{1\ldots n}.
\end{equation}Here, $S_k$ is the soft factor \eqref{eik} corresponding to the $k$-th photon and $A$ is the amplitude of the base process with all photons removed.
The number of terms in \eqref{soft_guess} is given by the number of photons in $A_n$. While the above function exhibits, by construction, the correct leading soft behavior for all photons,
it does not generically yield the correct amplitude.

An exceptional case where \eqref{soft_guess} yields valid results is for MHV amplitudes. In fact, recursively applying the above formula onto itself we strip off all photons in each term of \eqref{soft_guess} leading to
\begin{equation}\label{sftFctFrm}
    A_n = A \prod_{k} S_k,
\end{equation}
which is just the general soft factor form of MHV amplitudes discussed previously.

\begin{figure}[ht]
	\centering
	\includegraphics[width=0.2\linewidth]{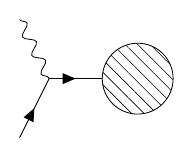}
	\caption{Schematic diagram which contributes to IR divergence of an amplitude. }
	\label{fig:sp}
\end{figure}

\begin{figure}[ht]
    \centering
    \includegraphics[width=0.32\linewidth]{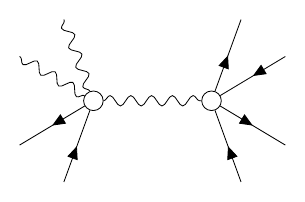}
\caption{Schematic factorization channel that will appear in 6 fermion recursion, where the right amplitude will generically be $\overline{\text{MHV}}$ or NMHV.}
    \label{6f_fact}
\end{figure}

The reason that MHV amplitudes admit such a simple soft factor form stems from the absence of three and higher particle poles. This can be directly connected with the Weinberg soft photon theorem. There, soft divergences only occur in Feynman diagrams when the soft photon is emitted from an external fermion line as shown in Figure \ref{fig:sp}.
These divergences are exactly captured by the leading soft factor $S_k$ and arise because of the presence of two particle poles in these diagrams. On the other hand, photon emissions from internal lines of Feynman diagrams involve multi-particle poles and are sub-leading in the soft limit. Therefore, they are not encoded in $S_k$. However, for MHV amplitudes these sub-leading contributions are absent and the amplitude can be fully determined in terms of soft factors $S_k$ as we showed. However, if our construction \eqref{soft_guess} can be consistently supplemented with information about higher particle poles, the amplitudes can be uniquely fixed\cite{Rodina2018}. An example of a factorization channel missed by the soft factors construction is shown in Figure \ref{6f_fact}.

\subsection{Graviton scattering}

The soft factor form of MHV amplitudes discussed so far derives from a detailed knowledge of their soft photon behavior.
Another theory whose soft properties have been extensively studied is gravity. There, soft graviton theorems analogous to those for photons (cf. \eqref{photSftThm}) can be used to derive soft factors
\begin{equation}\label{g_soft}
    S_k = \underbrace{\kappa \sum_{a=1}^n \frac{\spag{xa}\spag{ya}\spsq{ka}}{\spag{xk}\spag{yk}\spag{ak}}}_{S_k^{(0)}} + \underbrace{\frac{\kappa}{2}  \sum_{a=1}^n \frac{\spsq{ak}}{\spag{ak}}\left( \frac{\spag{ax}}{\spag{kx}} + \frac{\spag{ay}}{\spag{ky}}\right) \spsq{k \partial_a}}_{S_k^{(1)}} + \underbrace{\frac{\kappa}{2}  \sum_{a=1}^n \frac{\spsq{ak}}{\spag{ak}} \spsq{k \partial_a}^2}_{S_k^{(2)}},
\end{equation}
corresponding to the emission of a soft, positive helicity graviton ($++$) with momentum $k$. In the above, $x,y$ are arbitrary reference spinors as \eqref{g_soft} is not manifestly gauge-invariant. The explicit form of the subleading contributions $S_k^{(1)},S_k^{(2)}$ was only found recently \cite{Bern2014,Broedel2014}.

Some all-multiplicity formulae for MHV graviton amplitudes are already known in the literature, e.g. \cite{Nguyen2010,Hodges2013}, which however do not make primary use of the amplitudes soft behavior.
Our approach differs as we wish to obtain amplitudes exclusively by applying soft theorems, without assuming locality or unitarity, which might lead to simpler recursive relations, as was the case in QED.

Given the known soft structure of gravity, an immediate question to ask if MHV amplitudes involving gravitons admit a soft factor form similar to those for photons in \eqref{sftFctFrm}. Indeed, from explicit calculations we find that a slightly weaker soft factor structure of the form
\begin{equation}\label{sftGrvFrm}
    M_n(h^{--}h^{--}h^{++}\hspace{-4mm}\ldots h^{++}) = S_n M_{n-1}(h^{--}h^{--}h^{++}\hspace{-4mm}\ldots \cancel{h^{++}}),
\end{equation}
is preserved for pure graviton amplitudes at least up $n=8$.
Concretely we find that the following relations hold
\begin{align}\label{grav}
    \begin{split}
     &M_4 \left(h^{--} h^{--} h^{++} h^{++}\right) = S_4^{(0)} M_3\left(h^{--} h^{--} h^{++}\right), \\
     &M_{5,6,7}\left(h^{--} h^{--} h^{++}\hspace{-4mm}\ldots h^{++}\right)  = \left(S_{5,6,7}^{(0)} + S_{5,6,7}^{(1)}\right) M_{4,5,6}\left(h^{--} h^{--} h^{++} \hspace{-4mm}\ldots h^{++}\right), \\
     &M_{8}\left(h^{--} h^{--} h^{++}\hspace{-4mm}\ldots h^{++}\right) =\left( S_{8}^{(0)}  +S_{8}^{(1)}+S_{8}^{(2)} \right)M_{7}\left(h^{--} h^{--} h^{++}\hspace{-4mm}\ldots h^{++}\right).
    \end{split}
\end{align}

There are two properties of equalities \eqref{grav} that we should point out. First, they can be extended to various other particles using the $\mathcal{N} = 8$ supergravity component amplitudes as the MHV superamplitudes satisfy these relations as well. Secondly, there is a key difference for gravity amplitudes when compared to MHV amplitudes involving photons studied previously. Iteratively applying the soft factor relation \eqref{sftGrvFrm} onto itself does not produce valid amplitudes, i.e. $M_8 \neq S_8 S_7 S_6 S_5 S_4 M_3$ does not produce the correct eight-point amplitude. The reason is again an issue with local momentum conservation as discussed below \eqref{FullFermion}, where different choices of lower point amplitude to which the soft factors are applied lead to different results. The exact forms for which \eqref{grav} works can be obtained from the recursion relations and code found in \cite{Bourjaily2023}. 

However, when trying to compute the nine-graviton amplitude $M_9$ by analogy to \eqref{grav}, we were unable to find a suitable form of the eight-point amplitude that would produce the correct result. It is curious that the pattern of \eqref{grav} seems to break specifically at $n=9$. Currently, a deeper understanding of the exact reason for this obstruction is lacking. 
Nevertheless, we should emphasize that the space of inequivalent eight-point functions which agree with $M_8$ on momentum conservation is huge. As such, it may still be worthwhile to look for particular functional forms of amplitudes for $n\ge 8$ that allow to extend the soft factor construction to MHV amplitudes of arbitrary multiplicity in gravity.

\section{Conclusion}

In this paper we have presented a simple derivations of compact formulae for MHV amplitudes in various theories of massless quantum electrodynamics. We demonstrated that all MHV amplitudes can be written in a concise soft factor form. Starting from a base process amplitude involving only charged matter particles we can add an arbitrary number of photons simply by multiplying the corresponding soft factors. Thus, MHV amplitudes are determined up to any multiplicity by their respective low-energy behavior. Initial results were obtained by exposing the soft factor structure of scattering amplitudes with two and four charged matter particles in abelian gauge theories such as spinor and scalar electrodynamics (s)QED. 

Besides studying abelian photons we also analyzed MHV amplitudes in several related theories. At first we considered maximally supersymmetric $SU(2)$ Yang-Mills theory where the gauge boson corresponding to the diagonal generator $T^3$ was treated as a "non-abelian photon". Switching to a non-abelian theory not only permitted us to consistently deal with charged vectors but also allowed us to study MHV amplitudes with an arbitrary number of charged matter particles and expose their soft factor form. 
We then turned to abelian gauge theories with charged matter of arbitrary spins $>1$, and obtained inconsistent, locality-violating expressions for amplitudes in accordance with the Weinberg-Witten theorem. Nevertheless, the soft factor decomposition was still maintained even for these amplitudes. Beyond photons, the scattering of gravitons was also briefly studied. There, the soft factor structure was found to have limited applicability up to $n=8$ for MHV graviton scattering.

Although our analysis was limited to MHV amplitudes, it is tempting to ask if some of their simplicity can be carried over to NMHV amplitudes and beyond. 
As we argued, the soft limits of any N$^k$MHV amplitude are expected to be correctly captured by a soft factor form similar to the one presented here for MHV. Consequently, an interesting open question remains how to consistently incorporate corrections to the soft factor form to account for the multiparticle pole structure present in N$^k$MHV amplitudes.

\acknowledgments
 We thank Jaroslav Trnka and Johannes Henn for very useful discussions. This work is supported by GA\-\v{C}R 24-11722S and OP JAK CZ.02.01.01/00/22\_008/0004632.

\appendix

\section{Generalized Schouten identity}\label{apd1}

In this Appendix we derive the identity that allowed us to sum the results \eqref{4fermSumSimpl} and \eqref{ID2}. The sum obtained here is slightly more general as we focus on
\begin{equation}\label{genschout}
	\sum_{k \in I} \alpha_k \underbrace{\left<k\right| \otimes \ldots \otimes \left<k\right|}_{|I|-1 \text{ times}}= \frac{\overbrace{\left<1\right| \otimes \ldots  \otimes \left<1\right|}^{|I|-1 \text{ times}}}{\prod_{k \in I} \spag{1k}}.
\end{equation} 
Where $ I $ is some set of indices, $ |I| $ is its size, $ 1 $ is not its element and $\alpha_k$ are numbers. The equality \eqref{genschout} reduces to the Schouten identity for $|I| = 2 $ and can thus be seen as its generalization.

First, let us argue that the left side forms a basis in a space of symmetric tensors. The dimension of space of symmetric tensors $ S_k $ of rank $ k $ over a $ d $-dimensional vector space is 
\begin{equation}
	\text{dim} \: S_k = \begin{pmatrix}
		k + d -1 \\
        k
	\end{pmatrix} = |I|.
\end{equation}
In our case, we have $ k=|I|-1, d= 2 $. So, the size of our set is equal to the dimension of the relevant space. Linear independence can be proven by contradiction. Let us assume that it is linearly dependent, then there must exist some $ \beta_k $ such that
\begin{equation}\label{beta}
		\sum_{k \in I} \beta_k \left<k\right| \otimes \ldots \otimes \left<k\right| = 0.
\end{equation}
In addition, let us assume generic kinematics or that $ p_j \cdot p_i \neq 0$, for all $ i,j \in I $. Then we contract all spinor indices in \eqref{beta} with all but one elements of $ I $, let us call that element $ l $ and we obtain 
\begin{equation}
	\beta_l \prod_{k \neq l} \spag{kl}=0.
\end{equation}
But since we are working with generic kinematics, none of the spinor brackets can be zero, and therefore we can divide by them and are left with $ \beta_l = 0 $. However, this can be done for all possible labels and therefore this set is linearly independent. Hence, the left side of \eqref{genschout} forms a basis.

Now, let us derive the coefficients $ \alpha $. Analogously to the previous step we contract all free spinor indices with all but one elements of $ I $ and are left with
\begin{equation}
	\alpha_l \prod_{j\neq l} \spag{jl} = \frac{1}{\spag{1l}} \implies \alpha_l = \frac{1}{\spag{1l}\prod_{j\neq l}{\spag{jl}}}.
\end{equation}
All that remains is to return these coefficients to the original expression and contract them with other external spinors to obtain equalities utilized in this text.

\section{Validation of employed shifts}\label{apd2}

In this Appendix, the behavior of Feynman diagrams under shifts employed in Section \ref{QED} is analyzed to establish $ \lim\limits_{z \to \infty} A(z)  = 0$ as a sufficient condition for the calculations performed there to be valid. The sources of large $z$ scaling are propagators, wavefunctions and vertices. The boson and scalar propagators scale as $O\left(1/z\right)$, whereas the fermionic is flat $O(1)$. Shifted wavefunctions behave according to their helicity and so they will be mentioned independently. Finally, the only vertices that carry momentum and consequently $z$ dependence are three-point vertices in scalar QED.

\subsection{Spinor QED}

At first we study the process with four fermions and all plus helicity photons. There is only a single topology of Feynman diagrams which schematically yields
\begin{equation}
		\vcenter{\includegraphics[width=0.25\linewidth]{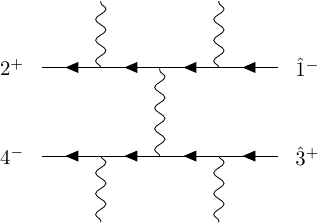}}\hspace{-10.5cm} \propto \left[2\right| \Gamma^\mu\left|1\right>\frac{1}{P^2} \left<4\right| \Gamma'_\mu\left|3\right].
\end{equation} Where $ \Gamma, \Gamma' $ are Dirac matrices built out of QED vertices and fermion propagators. The shift we employ is \eqref{shf}. However all of these diagrams have only one source of $z$ behavior due to the photon propagator and thus $O(1/z)$ scaling.

Changing our focus to the process with swapped fermion helicities, Feynman diagrams yield
\begin{equation}
	\vcenter{\includegraphics[width=0.25\linewidth]{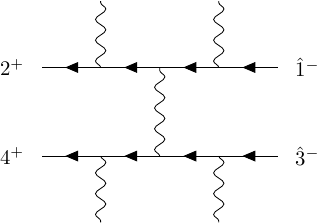}}\hspace{-10.5cm} \propto \left[2\right| \Gamma_\mu\left|1\right>\frac{1}{P^2} \left[4\right| \Gamma'_\mu\left|\hat{3}\right>.
\end{equation}
Here we can use the freedom in the definition of photon polarization vectors to take their reference spinor $\xi = p_1$, which eliminates $z$ scaling from all diagrams in which at least one photon is connected to shifted fermion lines are present. Thus the only potentially dangerous diagrams are
\begin{equation}
		\vcenter{\includegraphics[width=0.25\linewidth]{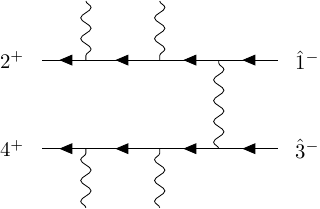}}\hspace{-10.5cm} \propto  \sum_\lambda\left[2\right|\Gamma\left|\lambda\right>\left[\lambda
		\right|\gamma^\mu\left|1\right>  \frac{1}{P^2}  \sum_\chi\left[4\right|\Gamma'\left|\chi\right>\left[\chi
		\right|\gamma_\mu\left|\hat{3}\right>.
\end{equation}
However, this expression is proportional to $ \spag{\hat{3}1} = \spag{31} $ and thus the $z$ dependence of wavefunctions is canceled. Consequently, all Feynman diagrams in our special choice of gauge are $ O\left(1/z\right)$, which originates from the photon propagator. That is exactly what we wanted to demonstrate.

\subsection{Scalar QED}

In the case of scalar QED wavefunctions scale trivially and the $z$ dependence of amplitudes only comes from three-point vertices. We will again exploit a specific choice of gauge to eliminate this scaling.

First, let us focus on a process with two charged scalars. The verification of this shift was done already in \cite{Badger2008} in a theory with massive charged scalars. Nevertheless, let us quickly derive this result for the sake of completeness. Zooming into a three-point vertex yields
\begin{equation}
		\vcenter{\includegraphics[width=0.25\linewidth]{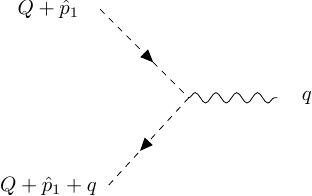}}\hspace{-10.5cm} \propto \left(Q+\hat{p}_1\right) \cdot \varepsilon(q,+) \propto \mix{\xi}{\hat{1}}{q} + O(1)
\end{equation}
By choosing $ \xi = p_2 $ all scaling is removed from not only this but all three-point vertices with a plus helicity photon. An analogous procedure can be performed for the minus helicity photon. Consequently, we find that there is no nontrivial scaling, which validates our shift. The previous argument fails at four-point, which can be constructed using a two-photon shift.

Now we turn to a process with four charged scalars and all plus helicity photons under the shift \eqref{SShift}. Once again, the only dangerous parts are the three-point vertices through which the shifted momentum flows. If there is an external photon in this vertex, the $ z $ scaling can be eliminated by proper gauge choice. Hence, we only have to focus on the cases where this argument fails. 

Focusing on a shifted photon vertex we find
\begin{equation}
	\vcenter{\includegraphics[width=0.25\linewidth]{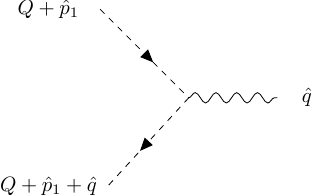}}\hspace{-10cm} \propto \left(Q+\hat{p}_1\right) \cdot \varepsilon(\hat{q},+) =\frac{\mix{\xi}{Q+\hat{p}_1}{q}}{\spag{\xi \hat{q}}}.
\end{equation}
But such an expression scales as $ 1/z $ as the shifted part vanishes. In addition, any off-shell photon has to be accompanied by at least one shifted scalar or photon propagator, as well as the polarization vector of the shifted photon. Together, they give rise to $1/z$ scaling. Thus, any Feynman diagram is $O\left(1/z\right)$ in the large $z$ limit. To finish, let us restate that this shift can be used from five-point onward as at least one photon has to be present.

Naturally, these arguments can be combined to validate shifts in theory with both charged scalars and fermions. Or be specifically used to construct \eqref{mix}.

\section{Seven-point example of four fermion formula}\label{7f}
Here we present the seven-point MHV amplitude for four charged fermions. This result more clearly illustrates the pattern that we extend into the ansatz \eqref{RES1}.

\begin{multline}
	A\left(f^-\Bar{f}^+f^+\Bar{f}^-\gamma^+ \gamma^+ \gamma^+\right) = \spag{14}^2\left(  \frac{\spag{12}^2}{\spag{34}\spag{15}\spag{25}\spag{16}\spag{26}\spag{17}\spag{27}} \right.\\
    +\frac{\spag{12}}{\spag{15}\spag{25}\spag{16}\spag{26}\spag{37}\spag{47}}
    +\frac{\spag{12}}{\spag{15}\spag{25}\spag{36}\spag{46}\spag{17}\spag{27}}
    +\frac{\spag{12}}{\spag{35}\spag{45}\spag{16}\spag{26}\spag{17}\spag{27}}\\
    +\frac{\spag{34}}{\spag{15}\spag{25}\spag{36}\spag{46}\spag{37}\spag{47}} 
     +\frac{\spag{34}}{\spag{35}\spag{45}\spag{16}\spag{26}\spag{37}\spag{47}}
    +\frac{\spag{34}}{\spag{35}\spag{45}\spag{36}\spag{26}\spag{17}\spag{27}}\\\left.
    +\frac{\spag{34}^2}{\spag{12}\spag{35}\spag{45}\spag{36}\spag{46}\spag{37}\spag{47}}\right).
\end{multline}

\section{All amplitudes computed in this text}

In this Appendix, we list all amplitudes computed throughout this text for the reader's convenience. The definition of eikonal factors in this text is
\begin{equation}
	S_k^{(m)} = \frac{1}{\sqrt{2}} \!\sum_{i=1}^{m}q_l \frac{\epsilon\left(k , +\right) \!\cdot p_i}{k \cdot p_i} = \sum_{\substack{j=1\\\text{odd}}}^{m-1}\frac{\spag{j\,j{+}1}}{\spag{jk}\spag{j{+}1\,k}}.
\end{equation}

\subsection*{Spinor QED}

\begin{align}\label{}
    \begin{split}
         	A\left(f^{h_1} \bar{f}^{h_2} \gamma^+ \ldots \gamma^+\right) = 	A\left(f^{h_1} \bar{f}^{h_2} \gamma^- \ldots \gamma^-\right)=0,\\
 	A_n\left(f^- \bar{f}^+ \gamma^- \gamma^+ \ldots \gamma^+\right) = \frac{\spag{13}^2}{\spag{12}} \prod_{k=4}^{n} S_k^{(2)}.
    \end{split}
\end{align}
The previous expressions were taken from \cite{Badger2010, Ozeren2005} and we list them for completeness.

\begin{align}\label{}
    \begin{split}
        A\!\left( f^-\bar{f}^+ f^+ \bar{f}^-\gamma^+\ldots \gamma^+\right) &=\frac{\spag{14}^2}{\spag{12}\spag{34}}\prod_{k=5}^{n}\! S_k^{(4)},\\
        A\!\left( f^-\bar{f}^+ f^-\bar{f}^+\gamma^+\ldots \gamma^+\right) &=\frac{\spag{13}^3\spag{24}}{\spag{12}\spag{23}\spag	{34}\spag{41}}\prod_{k=5}^{n}\! S_k^{(4)}.
    \end{split}
\end{align}

\subsection*{Scalar QED}

\begin{equation}
    		A\!\left(\phi \phi^*  \gamma^+ \ldots \gamma^+ \right) = A\!\left(\phi \phi^*  \gamma^- \ldots \gamma^- \right) = 0
\end{equation}
The previous amplitudes can be found in \cite{Badger2008}.
\begin{equation}
   A\!\left( \phi \phi^* \gamma^- \gamma^+ \ldots \gamma^+ \right) \frac{\spag{13}\spag{23}}{\spag{12}} \prod_{k=1}^n S_k^{(2)}.
\end{equation}

\begin{equation}
        A\!\left(\phi \phi^* \phi \phi^*  \gamma^+ \ldots \gamma^+ \right) = \left( \frac{\spag{13}^2\spag{24}^2}{\spag{12}\spag{14}\spag{32}\spag{34}} + C \right) \prod_{k=5}^n S^{(4)}_k
\end{equation}

\subsection*{Scalar-Spinor QED}

\begin{equation}
    A\!\left( f^- \bar{f}^+ \phi \phi^* \gamma^+ \ldots \gamma^+ \right) = \frac{\spag{13}\spag{14}}{\spag{12}\spag{34}}\prod_{k=5}^n S^{(4)}_k
\end{equation}

\subsection*{\texorpdfstring{$\mathcal{N} = 2$}{ N = 2} supersymmetry}

\begin{equation}
    	\mathcal{A}\!\left(\Phi \overline{\Phi} V^- V^+ \ldots V^+\right)= \frac{\gdf{4}}{\spag{12}}\prod_{k=4}^nS_k^{(2)}
\end{equation}

\begin{equation}
    	\mathcal{A}\!\left(\Phi \overline{\Phi} \Phi \overline{\Phi} V^+ \ldots V^+\right)= \frac{\gdf{4}\spag{13}\spag{24}}{\spag{12}\spag{23}\spag{34}\spag{41}}\prod_{k=5}^n S_k^{(4)}
\end{equation}

\subsection*{\texorpdfstring{$\mathcal{N} = 4$}{ N = 4} supersymmetry}

\begin{equation}
    	\mathcal{A}^{\text{MHV}}\!\left(W \overline{W} V\ldots V\right) = \frac{\gdf{8}}{\spag{12}^2}\prod_{ k =3}^n S_k^{(2)}.
\end{equation}

\begin{equation}
    	\mathcal{A}^{\text{MHV}}\!\left(W \overline{W} W \overline{W} V\ldots V\right) = \frac{\gdf{8}}{\spag{12}\spag{14}\spag{32}\spag{34}} \prod_{ k =5}^n S_k^{(4)}
\end{equation}

\begin{equation}
        \mathcal{A}^{\text{MHV}}_n \big(\underbrace{W \overline{W} \ldots W \overline{W}}_{2m} \hspace{1mm} V \ldots V \big) =  \mathcal{A}^{\text{MHV}}_{2m} \big(W \overline{W} \ldots W \overline{W} \big) \prod_{k=2m+1}^n S^{(2m)}_k.
\end{equation}
Where we have used the standard expression for MHV superamplitudes in Yang-Mills theory
\begin{equation}
        \mathcal{A}^{\text{MHV}}_{2m} \!\left(W \overline{W} \ldots W \overline{W} \right) =\!\!\!\!\sum_{\pi \in S_{2m}/Z_{2m}}^{} \!\!\!\!\text{Tr}\!\left[T^{\pi(a_1)}\ldots T^{\pi(a_{2m})}\right] \,\mathcal{PT}\!\left[\pi (1)\ldots\pi (2m) \right],
\end{equation}

\begin{equation}
    \mathcal{PT}\left(1\ldots 2m \right) = \frac{\gdf{8}}{\spag{12}\spag{34}\ldots \spag{2k{-}1 \; 2m}\spag{2m  \; 1}}.
\end{equation}

\subsection*{Higher spin QED}
We stress that the expressions in this part are computed formally, as they do not correspond to any physically permissible theory.
\begin{equation}
        A\!\left( 1^h \bar{2}^{-h} \gamma^- \gamma^+ \ldots \gamma^+\right) = \frac{\spag{13}\spag{23}}{\spag{12}} \left(\!\frac{\spag{23}}{\spag{13}}\!\right)^{\!2h} \prod_{k=4}^n S_k^{(2)}
\end{equation}

\begin{equation}
    A\left(1^h \bar{2}^{-h} 3^s \bar{4}^{-s} \gamma^+ \ldots \gamma^+\right) = \frac{\spag{13}\spag{24}}{\spag{12}\spag{34}} \left(\! \frac{\spag{24}}{\spag{14}}\!\right)^{\!2h}\! \left(\! \frac{\spag{14}}{\spag{13}}\!\right)^{\!2s}\prod_{k=5}^n S_k^{(4)}
\end{equation}

\subsection*{Gravity}
The soft factors for gravitons are defined as 
\begin{equation}
    S_q = \underbrace{\kappa \sum_{a=1}^n \frac{\spag{xa}\spag{ya}\spsq{qa}}{\spag{xq}\spag{yq}\spag{aq}}}_{S^{(0)}} + \underbrace{\frac{\kappa}{2}  \sum_{a=1}^n \frac{\spsq{aq}}{\spag{aq}}\left( \frac{\spag{ax}}{\spag{qx}} + \frac{\spag{ay}}{\spag{qy}}\right) \spsq{q \partial_a}}_{S^{(1)}} + \underbrace{\frac{\kappa}{2}  \sum_{a=1}^n \frac{\spsq{aq}}{\spag{aq}} \spsq{q \partial_a}^2}_{S^{(2)}}.
\end{equation}
Using them it is possible to find the following equalities
\begin{align}
    \begin{split}
     &M_4 \left(h^{--} h^{--} h^{++} h^{++}\right) = S_4^{(0)} M_3\left(h^{--} h^{--} h^{++}\right), \\
     & M_{5,6,7}\left(h^{--} h^{--} h^{++}\hspace{-4mm}\ldots h^{++}\right)  = \left(S_{5,6,7}^{(0)} + S_{5,6,7}^{(1)}\right) M_{4,5,6}\left(h^{--} h^{--} h^{++} \hspace{-4mm}\ldots h^{++}\right), \\
     &M_{8}\left(h^{--} h^{--} h^{++}\hspace{-4mm}\ldots h^{++}\right) =\left( S_{8}^{(0)}  +S_{8}^{(1)}+S_{8}^{(2)} \right)M_{7}\left(h^{--} h^{--} h^{++}\hspace{-4mm}\ldots h^{++}\right).
    \end{split}
\end{align}
The amplitudes on the right side are those obtained by standard recursion relations. 

\bibliography{ref}

\end{document}